\documentclass[intlimits,twoside,a4paper]{article}
\usepackage{amsmath,amssymb}
\usepackage{graphicx}
\usepackage[T2A]{fontenc}
\usepackage[cp1251]{inputenc}
\usepackage{graphics}
\usepackage{crop}

\usepackage{cmpj}

\issue{2011}{14}{4}{43801}

\doinumber{10.5488/CMP.14.43801}

\title[Capacitance of a high density electrolyte]%
{Application of the SRISM approach  to the study \\ of the capacitance of the double layer  of a high density primitive model electrolyte}

\author[S. Woelki, D. Henderson]{S. Woelki\refaddr{1}, D. Henderson\refaddr{2}}

\authorcopyright{S. Woelki, D. Henderson, 2011}

\addresses{
\addr{1}Institute of Physical and Theoretical Chemistry, University of Regensburg, Germany
\addr{2}Department of Chemistry and Biochemistry, Brigham Young University, Provo UT 84602, USA}

\date{Received October 17, 2011, in final form November 8, 2011}

\begin{document}

\maketitle

\begin{abstract}

In this study the Singlet Reference Interaction Site Model (SRISM) is
employed to the study of the electrode charge dependence of the capacitance of a
planar electric double layer using the primitive model of the double layer
for a high density electrolyte that mimics an ionic liquid.
The ions are represented by
charged hard spheres and the electrode is a uniformly charged flat surface. The capacitance of this
model fluid is calculated with the SRISM approach with closures based on the hypernetted chain (HNC)
and Kovalenko-Hirata (KH) closures and compared with simulations.  As long as the magnitude of the
electrode charge is not too great, the HNC
closure shows the most promise.  The KH results are reasonably good for a high
density electrolyte but are poor when applied at low densities.
\pacs 82.45.Fk, 61.20Qg, 82.45.Gj
\end{abstract}

\section{Introduction}

In a recent paper~\cite{WBH}, two of us together with Bhuiyan (WBH) applied the Singlet Reference Interaction Site Model (SRISM), using both the hypernetted chain
(HNC) and Kovalenko-Hirata (KH) closures, to the study of density profiles of the double layer formed by a restricted primitive model (RPM) at low electrolyte densities and concluded that the HNC closure was preferable.   However, that study was an introductory one whose purpose was to apply and test our programs to the previously studied systems.

The RPM is a simple, but informative, model of an electrolyte that models the ions as charged hard spheres, all of the same diameter, in a dielectric continuum that might represent the solvent in the case of ions dissolved in a solvent or represent the environment of the ions or a diminished partial charges of the ions in the case of a high density system.  It has been employed in the study of double layers for aqueous or other solvents.  An electrical double layer (EDL) is formed
by the ions near a charged electrode because the charge of the electrode attracts a compensating layer of counterions.
This compensating layer may be a single layer or may consist of several layers, some of which may consist predominantly
of coions.

In this paper, we study the model electrolyte first introduced by Lamperski et al.~\cite{LOB} and
recently considered by Outhwaite et al.~\cite{OLB}.
This model is based on the RPM and consists of charged hard spheres whose diameter is $d=4$~\AA~and
whose dimensionless temperature is $T^*=\epsilon dk_\text{B}T/z^2e_\text{o}^2=0.8$,
where $k_\text{B}$ is the Boltzmann constant, $T$ is the
temperature of the system and $e_\text{o}$ is the magnitude of the elementary charge.  The ions are assumed to be
symmetric with charges $ze_\text{o}$ and $-ze_\text{o}$ and $\epsilon=78.5$ is the background dielectric constant of the
environment of the ions.
Lamperski et al. (LOB) examined the capacitance of this model system by means
of a simulation and found that the capacitance had a minimum at zero electrode charge at small electrolyte densities.
Such a minimum is typical of the Gouy-Chapman-Stern (GCS) theory~\cite{GOUY,CHAP,STERN}, that neglects ion
diameters, except for the distance of closest approach of the ions, and correlations
between the ions.  This parabolic-like minimum is seen in simple
low density electrolytes.  As the density is increased, this minimum becomes shallower but is always
present in the GCS theory.  By contrast, LOB found, for their simulations that used their
model system, that at higher densities the minimum in the capacitance fills and becomes a maximum. As pointed out by Kornyshev~\cite{KORN}, such a
maximum is typical of many ionic liquids that exhibit a bell shaped or dromedary camel-like capacitance
curve.  Kornyshev's theoretical observations have been supported by recent experimental work by Lockett et al.~\cite{LOCK} and Islam et al.~\cite{ISLAM1,ISLAM2}.  Indeed, the LOB model was not meant to represent any specific
system but does mimic an ionic liquid.   LOB also found that the modified Poisson-Boltzmann (MPB)
theory~\cite{MPB} exhibited this same behavior.  Since the MPB theory fails to converge at high electrode
charge, they did not examine the capacitance at high electrode charge.  Subsequently, Lamperski and
Henderson (LH)~\cite{LH} extended this study to a very large electrode charge and found in their simulations
that the capacitance
at low densities has a minimum at zero electrode charge but the capacitance  has a
maximum at moderate electrode charge.  This maximum is followed by a steadily decreasing capacitance.
The capacitance at low densities has the shape of the letter `m', or a bird with drooping wings, or a
bactrian camel. At large
electrode charge, the capacitance in the LH study is density dependent but is not strongly so.  Thus,
at all electrolyte densities, the capacitance decreases steadily at a high electrode charge.  This is
due to a thickening of the double layer because there is a limit to the amount of counterion charge
that can be accommodated near the electrode.  By contrast, the GCS theory neglects the ion size and
correlations and there is no limit to the amount of charge that can approach the electrode in this theory.

The purpose of this paper is to apply the SRISM theory of our previous paper to the LOB
system.  The present study is an interim stage in our work since SRISM is not restricted to spherical ions.
Indeed, our long-range intention is to study double layers that contain nonspherical ions but our plan
is to proceed in a step by step manner.

\section{Theory}

We reviewed the SRISM theory in detail in our previous paper ~\cite{WBH} and there is little need to repeat this discussion.  Two closures were examined, the
hypernetted chain (HNC) and Kovalenko-Hirata (KH) closures.  The KH approximation is a hybrid of the HNC approximation and the mean spherical
approximation (MSA).  These closures employ the Ornstein-Zernike (OZ) relations, given by equation~(2) of WBH, that connects the total and direct correlation functions.
The total correlation functions are the radial distribution functions, less their asymptotic values (unity). Thus,
\begin{equation}
h_{ij}(r)=g_{ij}(r)-1,
\end{equation}
where the $h_{ij}(r)$ are the total correlation functions and the $g_{ij}(r)$ are the radial distribution functions
that give the probability of finding ions of species $i$ at a distance $r$ from a central ion of species $j$.
The direct correlation functions, $c_{ij}(r)$, are defined by the OZ relations.  As the name implies, the total
correlation functions give a complete information about all the correlations between a pair of ions whereas the
direct correlation functions give the correlations of such a pair directly, but in the presence of the other ions.
Since the OZ relations are a definition of the direct correlation functions, they do not yield a
method of calculating any of these correlation functions. The OZ relations must
be supplemented by an approximate closure.  For example, for the spherical system considered here, the HNC closure is
\begin{equation}
h_{ij}(r)=c_{ij}(r)+\ln g_{ij}(r)+\beta u_{ij}(r),
\end{equation}
where $\beta=1/k_\text{B}T$ with $k_\text{B}$ being the Boltzmann constant and $u_{ij}(r)$ is the pair potential between a
pair of ions of species $i$ and $j$ and charge $z_ie_\text{o}$ and $z_je_\text{o}$.  The pair potential is given by
\begin{alignat}{3}
u_{ij} (r) & = \frac{z_i z_j e_{\rm o}^2}{\varepsilon r} & \qquad\qquad
& \text{for} \qquad r \geqslant d
\nonumber \\
u_{ij} (r) & = \infty & \qquad\qquad & \text{for} \qquad r < d.
\label{eq:ionionpot}
\end{alignat}

The KH closure is a hybrid of the HNC closure and the MSA closure, which is a linearized version of the HNC closure, and is given by
\addtocounter{equation}{1}
\begin{alignat}{3}
c_{ij} (r) = & - \beta u_{ij} (r)  & \qquad\qquad &\text{for} \qquad r \geqslant d
\tag{\theequation a} \\
h_{ij} (r) = & -1 & \qquad\qquad & \text{for} \qquad r < d.
\tag{\theequation b} \label{eq:MSA_close}
\end{alignat}
The second part of the above equation is merely a statement of the fact that the ions are impenetrable.  The first part is the approximation.
Specifically, the KH closure is
\addtocounter{equation}{1}
\begin{alignat}{4}
c_{ij}(r) = & - \beta u_{ij}(r)  & \qquad & \text{for} \qquad  h_{ij}(r) -c_{ij}(r)- \beta u_{ij}(r) \geqslant  0
\tag{\theequation a} \\
c_{ij}(r)  = & \ h_{ij}(r) - \ln g_{ij}(r) -\beta u_{ij}(r) & \qquad & \text{for} \qquad
h_{ij}(r) -c_{ij}(r)-\beta u_{ij}(r) <  0
\tag{\theequation b} \\
h_{ij}(r)  = & -1 & \qquad & \text{for} \qquad r <  d.
\tag{\theequation c}
\label{eq:KH_close}
\end{alignat}
Again, the third part of the above equation is only a statement of the fact that the ions are impenetrable.

Our procedure is to calculate the pair correlation functions of the bulk electrolyte by solving the
bulk OZ equations with the chosen closure and then using these results as an input to solve the surface OZ equations
for the singlet surface correlation functions.  The ion-surface interaction potentials are
\begin{alignat}{3}
u_{\pm} (x) = & -\frac{4\pi z_\pm e_\text{o} \sigma}{\varepsilon}x  & \qquad\qquad & \text{for} \qquad x \geqslant d/2
\nonumber \\
u_{\pm} (x) = & \, \infty &\qquad\qquad & \text{for} \qquad x < d/2,
\label{eq:wallionpot}
\end{alignat}
where $x$ is the normal distance of the center of an ion from the electrode and $\sigma$ is the electrode charge density.  In some applications there is an advantage in using a separate closure for
the computation of the bulk electrolyte correlation functions.  This occurs when the MSA is used for a
bulk electrolyte with spherical ions because analytical results can be used.  We will not do this here
and instead use the same closure for the bulk and interfacial parts of our calculation.

Once the interfacial correlation functions have been determined, the interfacial charge density and electric
potential drop across the double layer can be determined by integration. For example,
\begin{equation}
\sigma = - 4\pi e_\text{o} \rho \int_{x=0}^{\infty} \left[ h_+ \left(x\right) - h_- \left(x\right)  \right] \text{d}x
\label{eq:sigma}
\end{equation}
and
\begin{equation}
\phi = -\frac{4\pi e_\text{o}\rho}{\varepsilon} \rho
\int_{x=0}^{\infty} \left[ h_+ \left(x\right) - h_- \left(x\right)  \right] x\text{d}x.
\label{eq:phi}
\end{equation}
We specify the charge density $\sigma$ as the input variable so equation~(\ref{eq:sigma}) is merely a check. Once $\phi$ as a function
of $\sigma$ has been determined, the integral capacitance,
\begin{equation}
C_{\mathrm i}=\frac{\sigma}{\phi}\, ,
\end{equation}
and differential capacitance,
\begin{equation}
C_{\mathrm d}=\frac{\partial \sigma}{\partial\phi}\, ,
\end{equation}
can be obtained. We use the dimensionless electrode charge density, $\sigma^*=\sigma d^2/e_\text{o}$ and electrode potential, $\phi^*=\beta e_\text{o}\phi$.  Following LH, we report values for the dimensionless differential capacitance, $C_{\mathrm d}d$.

\section{Results}

\begin{figure}[!h]
\vspace{1cm}
\centerline{\includegraphics[width=0.5\textwidth]{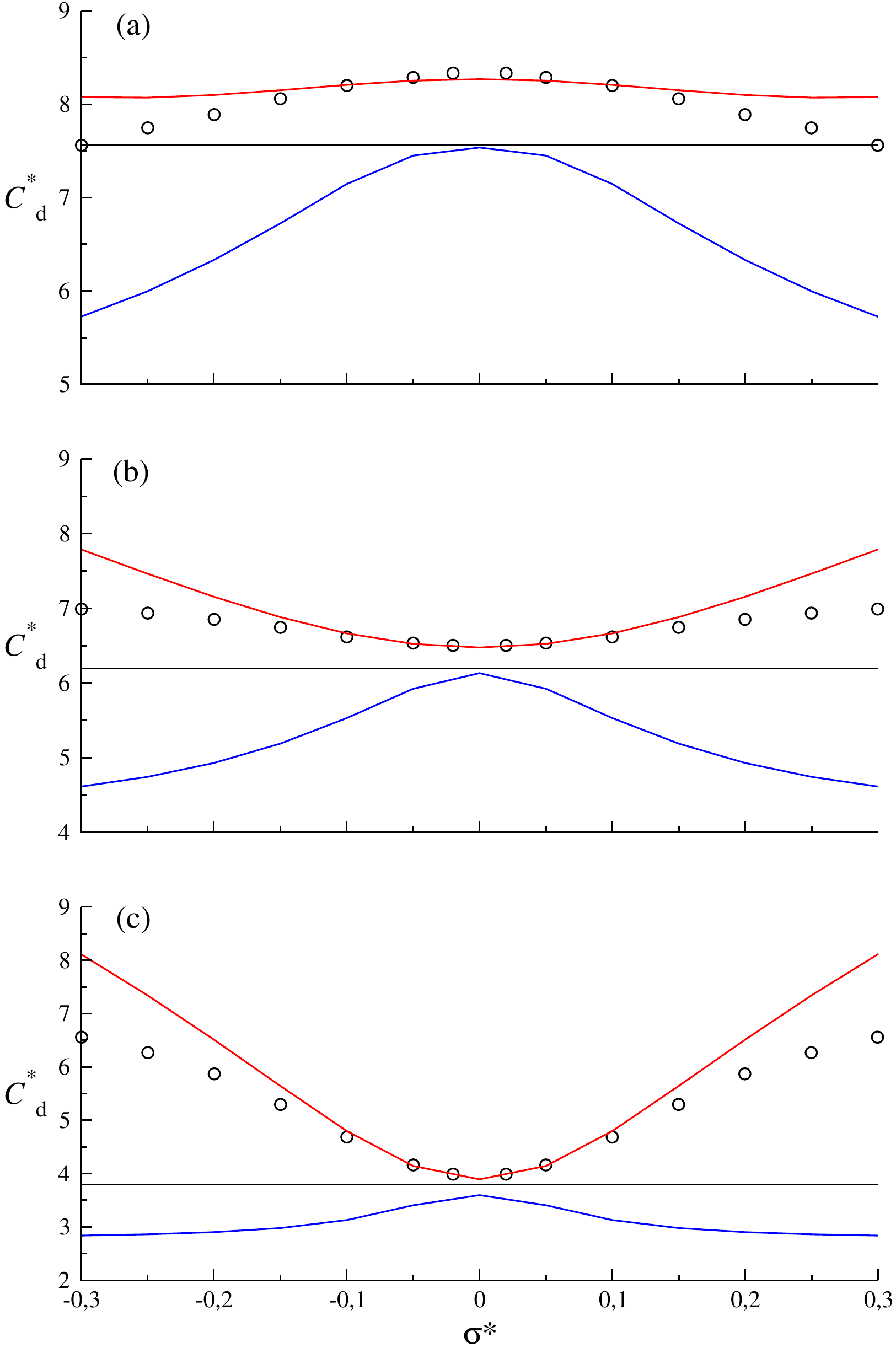}}
\caption{(Color online) The reduced differential capacitance, $C_{\mathrm d}^*=C_\text{d}d$, of the electrical double layer as a function of the
reduced surface charge density, $\sigma^*$ for $d=4$~\AA \ at the reduced temperature $T^* = 0.8$ and
the reduced electrolyte densities (a) $\rho^* = 0.24$, (b) $\rho^* = 0.14$ and (c) $\rho^* = 0.04$. The circles are the simulation results from reference~\cite{LH} and the black (middle), red (upper) and blue (lower) lines show the corresponding MSA/MSA1, HNC/HNC1 and
KH/KH1 results, respectively. \label{fig1}}
\end{figure}

\begin{figure}[!t]
\vspace{0.75cm}
\centerline{\includegraphics[width=0.5\textwidth]{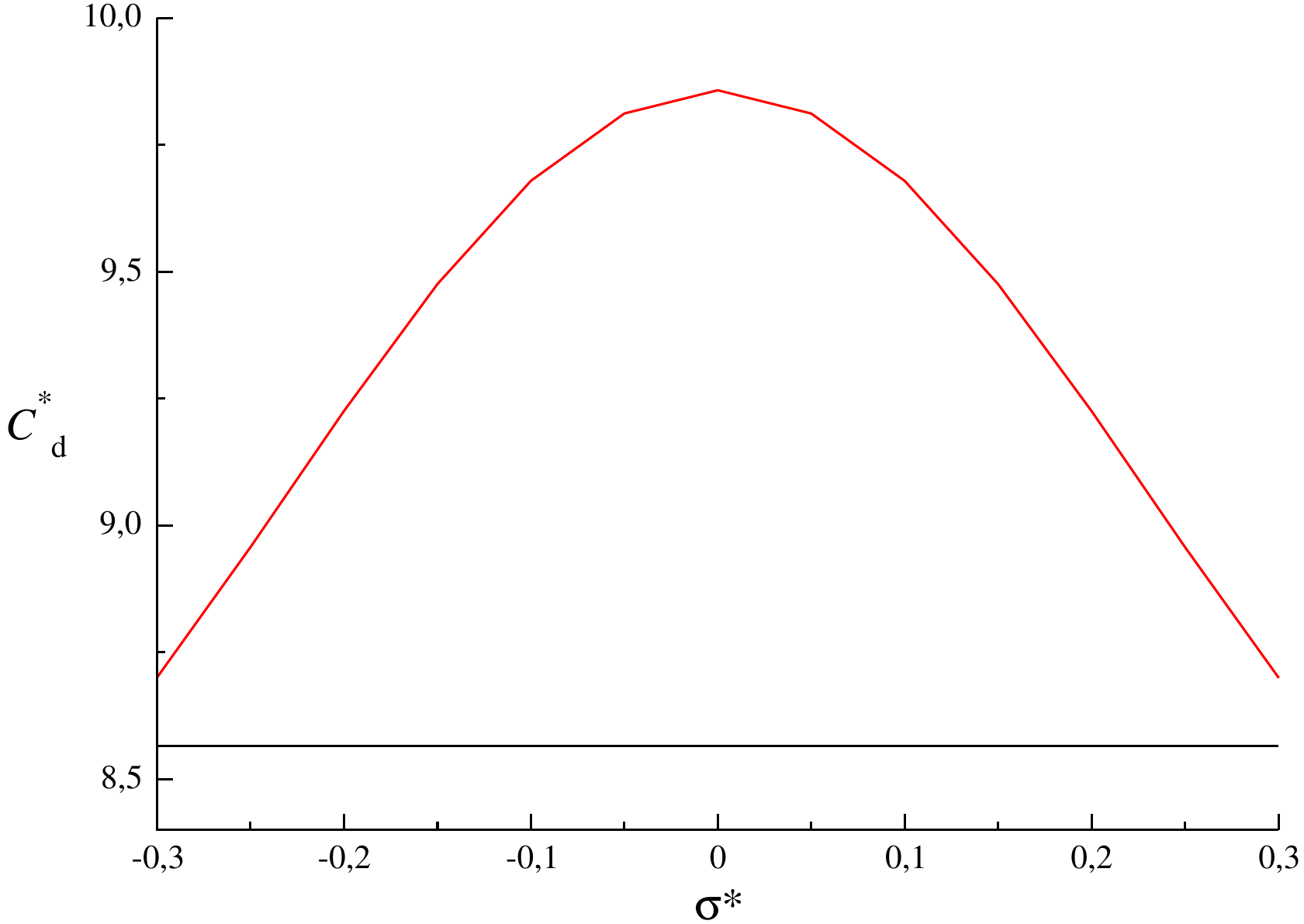}}
\caption{The reduced differential capacitance, $C_{\mathrm d}^*=C_\text{d}d$, of the electrical double layer as a function of the
reduced surface charge density, $\sigma^*$ for $d=4$~\AA~ at the reduced temperature $T^* = 0.8$ and
the reduced electrolyte density $\rho^* = 0.34$. The black (lower) and red (upper) lines show the MSA/MSA1 and HNC/HNC1 results,
respectively. \label{fig2}}
\end{figure}

In figure~\ref{fig1} we show our results for differential capacitances for the three densities, $\rho^*=Nd^3/V=0.04,0.14,0.24$, considered by LH.
Additionally, we show our HNC results for a higher density $\rho^*=0.34$ in figure~\ref{fig2}. By MSA/MSA1 we mean that the MSA is used for the bulk
correlation functions and the singlet interfacial correlation functions, respectively.  The SRISM theories fail for large $\sigma^*$ so we
restrict our comparison to $\sigma^*$ between $-0.3$  and $0.3$. The MSA results are quite reasonable at very low electrode charge. This has already
been reported by some of us~\cite{HLOB}. However, because the MSA is a linear response theory, the MSA capacitances are independent of $\sigma^*$.
The KH results are very poor for the $\rho^*=0.04$ system. They are rather like the MSA results because they change
little with $\sigma^*$. However, the KH results improve with increasing density. Indeed, the KH closure correctly predicts a maximum
in the capacitance for $\rho^*=0.24$.  Further study is needed to determine the general utility
of the KH closure for high density (ionic liquid) systems. The HNC results are fairly reasonable for $\rho^*=0.04$. For the highest densities,
the HNC predicts only a very weak maximum at $\rho^*=0.24$. However, it is at least a maximum. As is seen in figure~\ref{fig2}, there is a clear maximum
in the HNC capacitance at the highest density. Simulation results for this high density are not available and probably unnecessary as the purpose of figure~\ref{fig2} is merely to demonstrate that the HNC theory does exhibit a maximum in the capacitance at a sufficiently high density.  The relevant point is that the HNC theory
predicts a maximum at high densities but it is delayed.

\section{Summary}

We have applied the SRISM approach to the study of the capacitance of the double layer formed by charged hard spheres, all being of the same
diameter and magnitude of charge, near a flat uniformly charged unpolarizable surface for a range of
densities and electrode charge.  Comparison with simulations shows that the KH is rather poor at low
electrolyte densities but improves as the density is increased.  The HNC is fairly good overall but
the appearance of a maximum in the capacitance is delayed and does not appear clearly until rather high densities.
The MSA is fairly good for a small electrode charge but is incapable of accounting for the electrode charge
dependence of the capacitance due to of the linearization inherent in the approximation.  None of the
theories considered here is capable of predicting the decreasing values of the capacitance at a large
electrode charge because they neglect electrode correlations and permit too many ions near the
electrode and predict a double layer that is too thin.

On the whole, the MPB results are somewhat better than those reported here.   However, since the MPB theory, at least in
its present incarnation, fails to converge when the electrode charge is large in magnitude, it is not
known whether the MPB theory predicts the continuing decrease in the capacitance at large
electrode charge.  Also, at present, the MPB theory provides results only for spherical ions.  The results presented here
are not unpromising.  The HNC approximation, at
least, seems useful as long as the electrode charge is not too large.  The KH closure may be useful at high
electrolyte densities although a further study is needed to determine whether this is indeed the case.  Finally,
the SRISM approach can be applied to nonspherical ions.

The recent work by Loth et al.~\cite{Loth} presents an alternative theory and some simulations for the capacitance of a high density fluid.

\ukrainianpart

\title{Застосування методу  SRISM до вивчення  електромісткості подвійного шару  високогустинної примітивної \\ моделі електроліту}

\author{С. Уоелкі\refaddr{1}, Д. Гендерсон \refaddr{2}}

\addresses{
\addr{1}I Інститут фізичної і теоретичної хімії, Університет м.~Регенсбурга, Німеччина
\addr{2} Факультет хімії і біохімії, Університет Брігема Янга, Прово, США}

%
%
\makeukrtitle

\begin{abstract}
\tolerance=3000%
В цьому дослідженні  модель синглетного виділеного взаємодіючого центра  (SRISM)  застосовується до вивчення
залежності електромісткості  плоского електричного подвійного шару від заряду елек\-трода, використовуючи примітивну модель
подвійного шару для високогустинного електроліту, що  моделює іонну рідину. Іони представляються як заряджені тверді
сфери, а електрод є однорідно зарядженою плоскою поверхнею. Місткість цього модельного плину розраховується методом \mbox{SRISM}
із замиканнями, що базуються на гіперланцюжковому (HNC) підході і замиканнях Коваленка-Гірати (KH) і порівнюється з симуляціями.
Якщо величина заряду електрода не є дуже великою,  HNC замикання виглядає найбільш перспективним.  KH результати є прийнятними для
електроліту при високих густинах, але гіршими при низьких густинах.
\keywords  синглетні інтегральні рівняння, іонна рідина, обмежена примітивна модель, електромісткість
\end{abstract}

\end{document}